\documentclass[journal=jpcbfk,manuscript=article,layout=onecolumn]{achemso}
\usepackage[version=3]{mhchem} 
\usepackage[T1]{fontenc}      
\usepackage{graphicx}
\usepackage{subfigure}
\usepackage{multirow}
\usepackage{hhline}
\usepackage{chemarr}
\usepackage{amssymb}
\usepackage{color}
\usepackage{xspace}
\usepackage{chngcntr}
\counterwithout{table}{section}
\usepackage{textcomp}
\usepackage{gensymb}
\usepackage{xr} 
\usepackage{pdfpages}
\usepackage{booktabs}
\usepackage{multirow}
\usepackage[table,xcdraw]{xcolor}
\usepackage{colortbl} 
\usepackage{bm}
\usepackage{float}
\usepackage{ulem}
\usepackage{float}
\usepackage{hyperref}

\usepackage{soul}
\usepackage{xcolor}
\newcommand{\br}{\mathbf{r}}

\newcommand{\bmu}{\bm{\mu}}
\newcommand{\bT}{\mathbf{T}}

\newcommand{\bE}{\mathbf{E}}
\newcommand{\balpha}{\bm{\alpha}}

\newcommand{\bQ}{\mathbf{Q}}
\newcommand{\bV}{\mathbf{V}}

\author{Narjes Ansari}\affiliation{Qubit Pharmaceuticals, 29 rue du Faubourg Saint Jacques, 75014 Paris, France}\email{narjesa@qubit-pharmaceuticals.com}
\author{Zhifeng Francis Jing}\affiliation{Qubit Pharmaceuticals, 31 Saint James Avenue, Suite 810, Boston, MA 02116, United States}
\author{Antoine Gagelin}\affiliation{Qubit Pharmaceuticals, 29 rue du Faubourg Saint Jacques, 75014 Paris, France}
\author{Florent Hédin}\affiliation{Qubit Pharmaceuticals, 29 rue du Faubourg Saint Jacques, 75014 Paris, France}
\author{Félix Aviat}\affiliation{Qubit Pharmaceuticals, 29 rue du Faubourg Saint Jacques, 75014 Paris, France}
\author{Jérôme Hénin}\affiliation{Laboratoire de Biochimie Théorique,  UPR 9080 CNRS, Université de Paris Cité, 75005 Paris, France}
\author{Jean-Philip Piquemal}\affiliation{Qubit Pharmaceuticals, 29 rue du Faubourg Saint Jacques, 75014 Paris, France}\altaffiliation{Laboratoire de Chimie Théorique, Sorbonne Université, UMR 7616 CNRS, 75005 Paris, France}
\author{Louis Lagardère}\affiliation{Qubit Pharmaceuticals, 29 rue du Faubourg Saint Jacques, 75014 Paris, France}\altaffiliation{Laboratoire de Chimie Théorique, Sorbonne Université, UMR 7616 CNRS, 75005 Paris, France}\email{louis.lagardere@sorbonne-universite.fr}

\title{Lambda-ABF-OPES: Faster Convergence with High Accuracy in Alchemical Free Energy Calculations}

\begin{document}

\maketitle

\clearpage

\begin{abstract}

Predicting the binding affinity between small molecules and target macromolecules while combining both speed and accuracy, is a cornerstone of modern computational drug discovery which is critical for accelerating therapeutic development. Despite recent progresses in molecular dynamics (MD) simulations, such as advanced polarizable force fields and enhanced-sampling techniques, estimating absolute binding free energies (ABFE) remains computationally challenging. To overcome these difficulties, we introduce a highly efficient, hybrid methodology that couples the Lambda-Adaptive Biasing Force (Lambda-ABF) scheme with On-the-fly Probability Enhanced Sampling (OPES). This approach achieves up to a nine-fold improvement in sampling efficiency and computational speed compared to the original Lambda-ABF when used in conjunction with the AMOEBA polarizable force field, yielding converged results at a fraction of the cost of standard techniques.

\end{abstract}


Accurately predicting the binding affinity between a small molecule (ligand) and its target (protein or RNA/DNA) is a fundamental tool in modern computational drug discovery, that can be used both at the hit discovery and at the lead optimization stages~\cite{de2008computational}. Indeed, it minimizes the cost of drug discovery by reducing the need to rely on experimental assays.

Recent advancements in molecular simulation, especially in molecular dynamics (MD), have greatly enhanced our ability to predict binding affinities. One of the most complex challenges in this area is the prediction of absolute binding free energies (ABFE)~\cite{schindler2020large}, which depend on two key factors: reliable force fields (FF) and efficient sampling \cite{bernardi2015enhanced,henin2022enhanced,mehdi2024enhanced}. Force fields must accurately capture complex and subtle molecular interactions—such as electrostatics (ELE), van der Waals (VDW), and solvation effects—while also accounting for factors like protonation state variations and conformational flexibility. Efficient sampling is also essential to explore the vast configurational space of ligand-protein interactions and is often limited by complex free energy barriers. 

In this context, the development of high-quality polarizable force fields \cite{gresh2007anisotropic,reviewpol,melcr2019accurate}, such as AMOEBA~\cite{ponder_current_2010,wu2012automation,shi_polarizable_2013,AMOEBA-nucleic}, and tools like Poltype~\cite{poltype2} for accurately parametrizing complex ligands, have significantly enhanced the accuracy of MD simulations~\cite{laury2018absolute,D1SC00145K,D1SC05892D,chung2023accurate,D4SC04364B,ansari2024targeting}. These progresses allow for more accurate modeling of ligand-protein interactions and a more reliable prediction of associated thermodynamic properties, such as the binding affinity\cite{ansari2024targeting}.

Still, MD simulations encounter sampling limitations, especially in biological systems such as ligand-protein complexes. Indeed, the exploration of the conformational space is often hindered by significant free-energy barriers, associated to time scales beyond the reach of unbiased MD simulations, making it challenging to sample all relevant states exhaustively in standard simulations.

To address these challenges, several enhanced sampling algorithms have been developed. Notably, Collective Variable (CV)-based importance-sampling techniques such as Umbrella Sampling (US)\cite{torrie1977nonphysical}, Adaptive Biasing Force (ABF)\cite{darve2001calculating,darve2008adaptive,comer2015adaptive}, MetaDynamics (MtD)\cite{laio2002escaping,barducci2011metadynamics}, its recent evolution, On-the-fly Probability Enhanced Sampling (OPES)\cite{Invernizzi2020,Invernizzi2022}, and Temperature-Accelerated Molecular Dynamics (TAMD)~\cite{maragliano2006temperature} have demonstrated significant success. These algorithms rely on the definition of CVs, reduced dimensions along which biasing forces, potentials, or probabilities are applied to facilitate sampling of the system's configuration space. Each enhanced-sampling method has its advantages and limitations~\cite{henin2022enhanced}. To overcome these challenges, hybrid methods have been developed to combine different techniques and mitigate their individual limitations\cite{fu2019taming}.

Another class of methods, called alchemical approaches, estimates free energy differences associated to unphysical ``alchemical'' changes of the molecular Hamiltonian~\cite{williams2018free} by scaling some key interactions with an alchemical parameter $\lambda\in [0,1]$. They have proven useful to compute solvation as well as binding free energies. Traditional techniques involve several simulations at fixed $\lambda$ values and the reconstruction of the free energy difference of interest through an estimator such as the Free Energy Perturbation (FEP)~\cite{zwanzig1954high} or the Thermodynamic Integration (TI)~\cite{straatsma1991multiconfiguration} one.

Alternatively, another approach called lambda-dynamics~\cite{kong1996lambda} was introduced, in which the coupling parameter $\lambda$ is treated as a dynamical variable through an extended Lagrangian/Hamiltonian scheme. Building upon this concept, we introduced a new alchemical method, Lambda-Adaptive Biasing Force (Lambda-ABF)~\cite{lagardere2024labf}, that leverages lambda dynamics in combination with multiple-walker ABF~\cite{minoukadeh2010potential}, enabling efficient sampling of $\lambda$ as a collective variable (CV). It has been shown to be robust and to improve sampling efficiency compared to standard fixed-lambda methods~\cite{lagardere2024labf}. Additionally, in the context of binding simulations, it leverages Distance-to-Bound-Configuration (DBC)~\cite{henin2018dbc} restraints to keep the ligand within the binding pocket, limiting its translational, rotational, and conformational fluctuations. Compared to fixed-lambda methods, Lambda-ABF not only reduces the computational cost but is also arguably simpler as it bypasses the definition of a $\lambda$ schedule and portable thanks to its implementation within the Colvars library~\cite{fiorin2013using,fiorin2024expanded}. 

ABF applies biasing forces along the transition coordinate to flatten the sampled free-energy landscape toward a uniform distribution. It is well understood mathematically~\cite{comer2015adaptive} and is associated to a local (unconstrained) TI free-energy estimator. A key limitation arises with barriers along other degrees of freedom which can lead to kinetic trapping in some regions and overall non-ergodic sampling and also in the diffusive regime when (local) convergence is reached.

To overcome this limitation, we combine Lambda-ABF with the exploratory version of the OPES method, known as OPES-Explore~\cite{Invernizzi2022} (OPES$_{e}$). 
This method is CV-based and aims at sampling a target probability distribution associated to lower barriers. It can be seen as an evolution of (well-tempered) MetaDynamics with an emphasis on exploration and is user-friendly as it requires only a few physically motivated parameters to be set. By applying a bias to the lambda CV within the Lambda-ABF-OPES method, we effectively integrate the strengths of both ABF and OPES$_{e}$.

Similarly to the (wt)meta-eABF method~\cite{fu2019taming}, the Lambda-ABF-OPES approach benefits from the combination of ABF and OPES$_{e}$. While ABF reduces the free energy barriers, OPES$_{e}$ fills the energy valleys by incorporating a history-dependent potential term. To our knowledge, this is the first instance of combining ABF with the OPES$_{e}$ method in an alchemical framework. This approach improves convergence speed, enhancing it by up to 9 times compared to the original Lambda-ABF method.

We applied this hybrid approach, utilizing the AMOEBA polarizable force field and DBC restraints, to calculate the absolute binding free energy of 11 diverse small drug-like molecule inhibitors binding to bromodomains (BRD4) as well as the Benzamidine-Trypsin complex, a well-studied macromolecular systems~\cite{gapsys2021accurate,karrenbrock2024absolute,aldeghi2016accurate,bhati2024equilibrium,ansari2022water}. Our results show that while maintaining high accuracy, with a mean absolute error of 0.90~kcal/mol for the BRD4 system, we achieved a significant improvement in convergence speed. For the Benzamidine-Trypsin complex, we also obtained very close agreement with the experimental value, with a notably low computational cost compared to, for example, the results of Ref.\citenum{ansari2022water}. This represents a notable advancement, particularly when using the more computationally demanding AMOEBA polarizable force field.

The close agreement between our computational predictions and experimental data demonstrates the robustness of our hybrid method, offering a promising approach for computational drug design.



In alchemical methods, absolute binding free energy calculations critically depend on a precise definition of the bound state and well-designed ligand restraints (both translational and, optionally, orientational) to ensure rapid convergence. In this study, we employed the DBC coordinate to restrain the ligand during Lambda-ABF-OPES simulations. It is defined as the RMSD of some ligand atoms (LA) for each frame, with the alignment performed relative to some atoms of the receptor’s binding (RB) site~\cite{henin2018dbc}. This approach captures positional, orientational, and also conformational deviations of the ligand in a single collective variable.

In the case of BRD4 complexes, to select the LA, we monitored the root mean square fluctuation (RMSF) of the ligand's heavy atoms over at least 100 ns of standard MD simulations. Further details on simulation setup can be found in Supporting Information. Atoms with an RMSF of less than 0.6~\AA~were selected for tight binders (Ligands 1-9), while those with an RMSF between 0.7 and 0.8~\AA~were chosen for weak binders (Ligands 10-11). For the Benzamidine-Trypsin complex, due to the small size of the ligand, all heavy atoms of the ligand were selected for the DBC. 

For the RB selection of BRD4 and Trypsin, we used the C$\alpha$ atoms of the protein within 6~\AA~of the ligand, with an RMSF below approximately 0.6~\AA. See Fig. S1 in the SI for the definition of DBC for the protein and each ligand. During plain MD of each ligand, the DBC was monitored using the Colvars library, and the DBC value within the 95\% interval of the distribution was selected as the DBC cutoff for the restraint in Lambda-ABF-OPES. This selection ensures that during the alchemical simulation, there is no biased artifact from the DBC restraint. The DBC cutoff for each ligand is listed in Table S1 SI. A flat-bottomed harmonic restraint was then applied to the DBC with a force constant of 100~$\text{kcal/mol/Å}^2$ above the DBC cutoff.


Details of the Lambda-ABF and OPES$_{e}$ methods can be found in the original publications~\cite{lagardere2024labf,Invernizzi2020,Invernizzi2022}. Here we provide a brief overview of their underlying theories. 

The Lambda-ABF algorithm~\cite{darve2001calculating,darve2008adaptive,lagardere2024labf} adaptively computes the derivative of the free energy associated with the parameter $\lambda$ using the thermodynamic integration (TI) formula~\cite{kirkwood1935statistical}. This estimate is applied as a force directly to the simulated system, guiding the dynamics. As a result, the sampling of $\lambda$ converges toward a uniform distribution, facilitating the crossing of free-energy barriers.

OPES$_{e}$~\cite{Invernizzi2022}, a collective-variable (CV)-based enhanced sampling technique, represents the latest advancement in the MetaDynamics family. It broadens the sampling of a system toward a target probability distribution, known as the well-tempered distribution. This approach employs Gaussian kernels to adaptively construct a bias potential, guiding the system's exploration while preserving thermodynamic relevance. Critical parameters, such as the Gaussian kernel width and bias factor, are pivotal in defining the sampling's scope and stability. Additionally, the barrier parameter (\( \Delta E \)) ensures efficient transitions between basins while preventing access to irrelevant high-energy states.

In this study, the ABFE is calculated by continuously ``alchemically''  decoupling the ligand from its environment, both in complex with the protein and separately in bulk solvent. The standard free energy of binding is then determined using a thermodynamic cycle~\cite{santiago2023computing}, which requires sampling of the alchemical Hamiltonians. We employed the Lambda-ABF approach as implemented in Tinker-HP/Colvars, in combination with OPES$_{e}$ provided by the Colvars library~\cite{fiorin2013using} version 2024-11-18. This integration provides user-friendly convergence estimation without requiring post-processing and allows seamless compatibility with other CV-based methods.

All simulations were performed at T = 300~K and P = 1~Atm using the BAOAB-RESPA integrator~\cite{lagardere_pushing_2019} with a 3~fs time step under the NPT ensemble, employing four walkers~\cite{lelievre2007computation} to separately decouple the VDW and ELE interactions. More precisely, the polarizabilities and permanent multipoles of the ligand are scaled down to 0 in the electrostatic legs (ELE), and the van der Waals interactions between the atoms of the ligand and all the other ones are scaled down to 0 (leveraging softcore interactions~\cite{beutler1994avoiding}) in the van der Waals legs (VDW). In the complex phase, 30 ns were simulated for the VDW leg and 5 ns for the ELE leg. In the solvent phase, both the VDW and ELE legs were simulated for 5 ns each.



As mentioned earlier, for OPES$_{e}$, the barrier parameter is a critical setting in the simulations and must be sufficient to ensure that the method accurately captures the energy landscape and transition state barriers. We conducted tests using OPES$_{e}$ alone with different barriers. The final $\Delta G$ values varied between runs (see SI Fig. S20), indicating that OPES$_{e}$ alone struggles to compute the binding free energies of interest. However, when combining OPES$_{e}$ with Lambda-ABF, the system is primarily driven by Lambda-ABF, which facilitates the crossing of energy barriers, and also benefits from the rapid convergence of the TI estimator. In this case, OPES$_{e}$ mainly serves to push the system out of local minima. Therefore, the barrier parameter does not need to be equal to the free energy differences associated to the simulation. Our tests showed that setting the bias threshold of OPES$_{e}$ to a maximum of 5~kcal/mol for both the ELE and VDW legs in the complex and solvent phases effectively accelerated the convergence across all ligands in the BRD4 and Trypsin systems, regardless of their respective energy barriers. For ligand 11 in complex with BRD4, a threshold of 2 kcal/mol was sufficient for proper acceleration. Although a barrier of 5 kcal/mol also works well, since this ligand is a weak binder, the lower threshold of 2 kcal/mol ensures better stability in convergence. An adaptive sigma is used to determine the Gaussian kernel width, and the frequency for kernel deposition was set to 300 steps, which corresponds to 9~ps with a time step of 3~fs.


The free energy cost associated with the release of the DBC restraint is computed in the gas phase through TI by progressively releasing it to a compatible harmonic distance restraint which can then be computed analytically \cite{lagardere2024labf,santiago2023computing,henin2018dbc}.

In the BRD4 simulations, we observed that the rotamers of Asn140 in the Apo state may adopt a conformation favorable in the Apo state but not in the Holo state (see Results and Discussion and SI for more details). This could lead to sampling the Holo state with an incorrect orientation of Asn140, potentially introducing artifacts. The restriction (through a restraint) to one of these rotamers as an Apo endpoint must be accounted for. Therefore, a positional restraint was applied to the heavy atoms of the Asn140 residue and its neighboring atoms (all heavy atoms of the backbone and those within 6 \AA \ of the ligand with a mild force constant of 2~$\text{kcal/mol/Å}^2$). This restraint was necessary to ensure that Asn140 and its neighboring atoms were properly sampled in both the Apo and Holo states, given the continuous switching of lambda between 0 (Apo) and 1 (Holo) in the Lambda-ABF-OPES simulations. Since both rotamers of Asn140 are equally favorable in the Apo state (see Fig.~\ref{fig:fig2}), an RTln(2) correction was added to the final computed $\Delta G$. In addition, to avoid artifacts during alchemical decoupling as described in SI, the C$\alpha$ atoms of BRD4 were restrained to their relaxed configuration.

As all TI-based technique, Lambda-ABF requires the computation of potential derivatives with respect to the alchemical parameter $\lambda$ which are not trivial to compute for many-body interactions such as polarization as present in the AMOEBA force field. In our previous work\cite{lagardere2024labf} we used a simple interpolation of polarization between the end states which gives immediately the associated derivatives as the difference of these, but this is associated with an increase of the computational cost because of the need to solve 2 polarization equations at each timestep. Recently, analytical derivatives associated to a simple scaling of the polarizabilities and multipoles have been introduced\cite{thiel2024constant}. Here we resorted to a more general formulation relying on the variational formulation of the many-body term and the Hellman-Feynman theorem: $E_{pol}(\br,\lambda)=E_{pol}(\br,\lambda,\bmu(\br,\lambda))$ and \[\frac{d E_{pol}}{d \lambda}=\frac{\partial E_{pol}}{\partial \lambda}+\frac{\partial E_{pol}}{\partial \bmu}\frac{\partial \bmu}{\partial \lambda}=\frac{\partial E_{pol}}{\partial \lambda}\] because of the minimum conditions on the induced dipoles. Note that this formulation still holds for other many-body terms with similar variational formulation as is the case for fluctuating charges\cite{lipparini2011polarizable}, continuum solvation models\cite{polcontinuum} or QM-MM\cite{loco2016qm,locoaccchem}. Further development of the gradients used in this work using Particle Mesh Ewald is given in Technical Appendix.


\begin{figure}[H]
	\centering
	\includegraphics[width=0.9\textwidth]{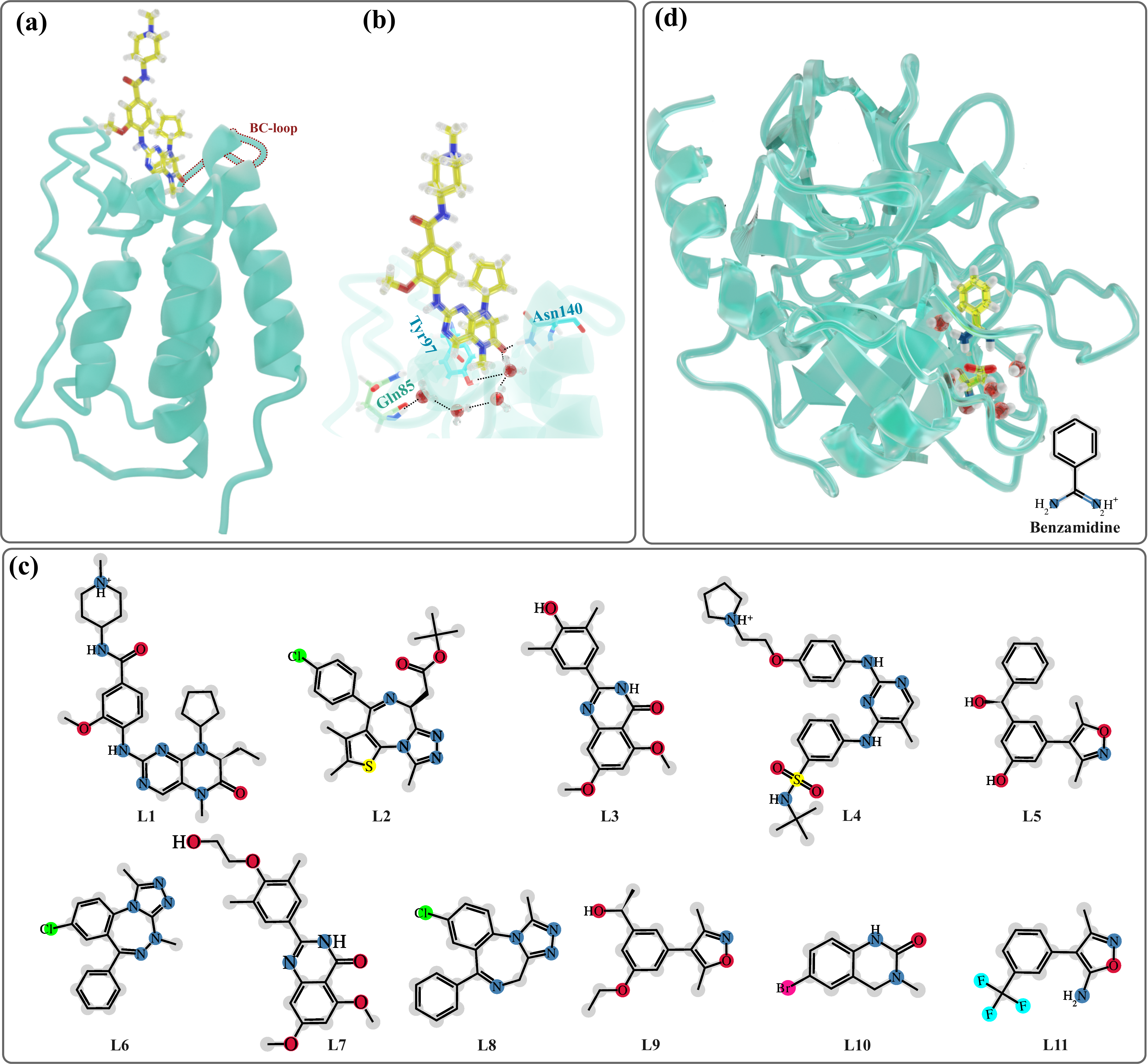}   
\caption{\textbf{Structural representation of the BRD4-ligand and Benzamidine-Trypsin complexes}.  (a) Cartoon representation of the BRD4 bromodomain structure in complex with ligand 1 (PDB ID: 4OGJ), with the BC-loop highlighted using a dashed marker.  (b) Binding mode of BRD4 with the ligand, highlighting key interacting residues Asn140 and Tyr97 in cyan. Crystallographically observed water molecules are shown as ball-and-stick representations, along with the hydrogen bond network bridging the ligand and protein via Tyr97 and Gln85. The protein is represented as a transparent cartoon.  (c) The 2D structures of the compounds analyzed in this study, labeled with Arabic numerals in descending order of binding affinity.  
(d) Cartoon representation of the Trypsin structure in complex with Benzamidine (PDB ID: 3ATL). The 2D structure of Benzamidine is also shown in the bottom-right corner of panel (d).} 

	\label{fig:fig1}
\end{figure}

In the following section, we will first provide a brief overview of the general binding modes of the 11 ligands in the BRD4 and Benzamidine-Trypsin system. We will then focus on the orientation of the Asn140 residue of BRD4, a key residue that directly interacts with the ligands, in both the Apo and Holo states, highlighting any significant differences. Finally, we will examine the binding affinities and correlate these findings with the experimental values.

As depicted in Fig.~\ref{fig:fig1}(c), for BRD4 the 11 ligands are large, flexible, drug-like molecules, some of which are charged (ligand 1 and 4). These characteristics make them an ideal set for evaluating the performance of the new method. All ligands target a common binding site. Asparagine (Asn140), located within the BC-loop binding pocket (Fig.~\ref{fig:fig1}(a)), is the most critical residue, which interacts directly with the ligands. The side chain of Asn140 consists of an amide group (\(-\text{CONH}_2\)) attached to a C$\beta$, which is itself connected to the C$\alpha$ of the backbone via rotatable bonds. 

The 100 ns plain MD simulations for all ligands (except ligand 11, for which 180 ns was run) show that, in the Holo state, only one rotamer is favorable due to the formation of salt bridges between the ligands and the Asn140 residue (Fig.~\ref{fig:fig1}(b)). However, in the Apo state, both rotamer states can be present. To investigate this, we calculated the free energy associated with the two rotamers in the Apo state using the OPES method (see SI for more details). Fig.~\ref{fig:fig2} shows that the two rotamers are equally favorable and that switching between them can happen naturally. 

In addition to Asn140, four conserved (polarizable) water molecules~\cite{aldeghi2018large} in the binding site of BRD4 also play a crucial role in stabilizing the ligands within the binding pocket, thereby enhancing binding affinities. As illustrated in Fig.~\ref{fig:fig1}(a), one of these water molecules forms a bridge between the Tyr97 residue and the ligand (except for ligand 4). To preserve this stabilization, the X-ray water molecules were retained in the binding pocket during system preparation, as they contribute to a hydrogen-bond network involving the three other conserved water molecules and the protein.

The Benzamidine-Trypsin complex (see Fig.~\ref{fig:fig1}(d)) serves as another ideal case to evaluate the performance of the Lambda-ABF-OPES on a different target.

Using the equilibrated structures for both targets, we carried out absolute binding free energy calculations employing the novel Lambda-ABF-OPES method. The correlation plot between experimental values and calculated results of BRD4 is shown in Fig.~\ref{fig:fig3}(a) and reported in Table~\ref{tab:EXP_Cal}.


\begin{figure}[H]
	\centering
	\includegraphics[width=0.8\textwidth]{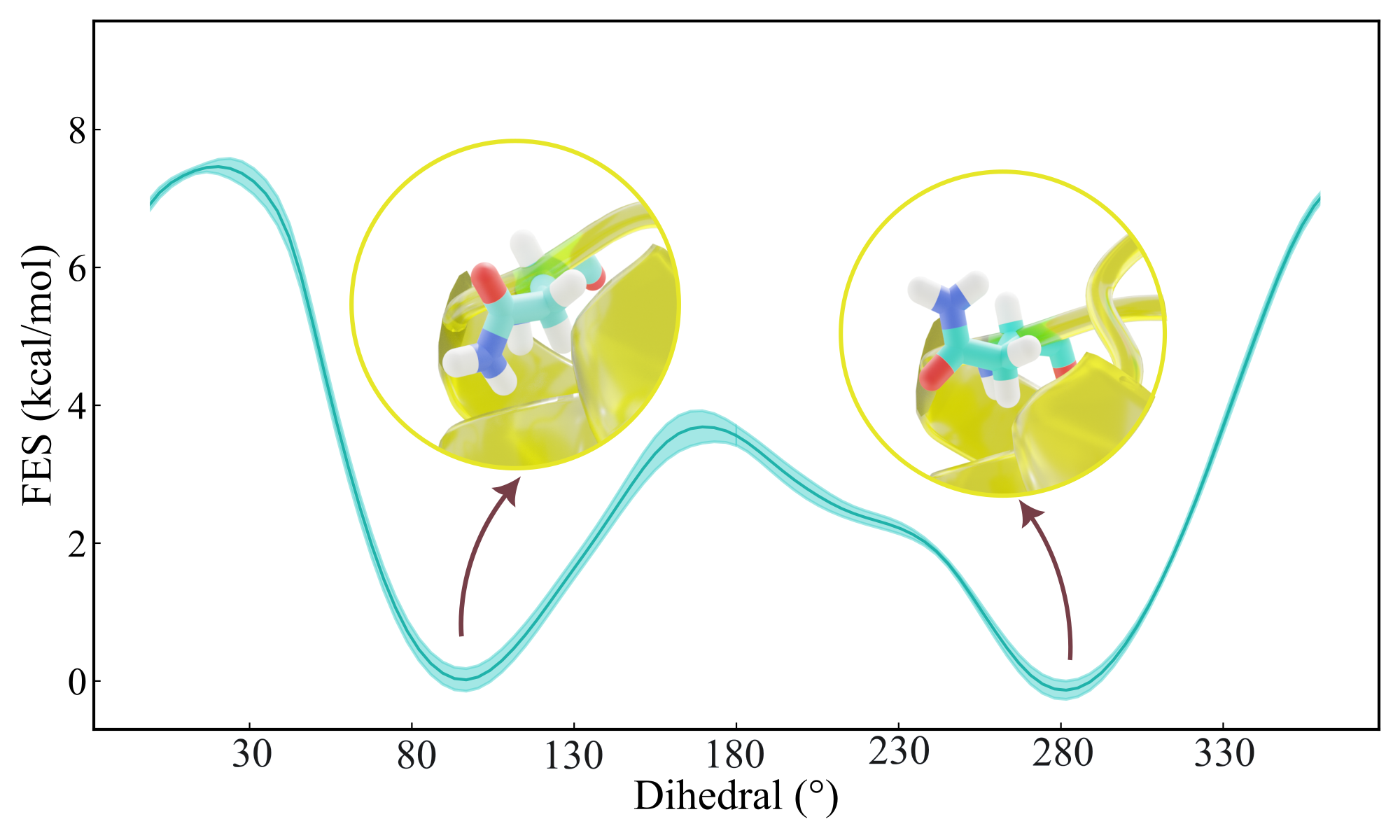}   
	\caption{\textbf{Free energy surfaces (FES) of the $\chi_{2}$ dihedral angle for the Asn140 residue} obtained from plain OPES simulations, starting the Apo structure (PDB ID: 4LYI). Two different rotamers of Asn140 corresponding to each minimum are represented. The protein is depicted in a cartoon representation in yellow, while Asn140 is shown as a stick representation. $\chi_{2}$ is the torsion angle between the nitrogen and the carbonyl carbon of the amide group. The transparent regions indicate the associated errors of the FES, calculated over block analysis.
}

	\label{fig:fig2}
\end{figure}


 We observe a strong correlation between the experimental and calculated results, with a Pearson's correlation coefficient (Pearson's $r$) of 0.81. The analysis yielded a root mean square error (RMSE) of 1.1~kcal/mol and a mean absolute error (MAE) of 0.9~kcal/mol. These results are consistent with experimental data (see Fig.~\ref{fig:fig3}, panel (a)) and align well with those reported in Ref.~\citenum{karrenbrock2024absolute} (see Fig.~\ref{fig:fig3}, panel (b)). For ligand 6, however, our obtained result aligned better with the experimental values. The key advantage of this method lies in its user-friendliness, reduced computational resource requirements, and rapid convergence. Compared to Lambda-ABF, the traditional fixed Lambda approach~\cite{aldeghi2016accurate}, and the newly developed CV-based approach~\cite{karrenbrock2024absolute}, this method offers significantly improved computational efficiency while maintaining high accuracy. These attributes make it particularly well-suited for large-scale or high-throughput applications.
 
For the Benzamidine-Trypsin complex, we obtained an absolute binding free energy of \(-6.2 \pm 0.65\) kcal/mol. This result is in excellent agreement with the experimental value of \(-6.3\) kcal/mol~\cite{talhout2003understanding,talhout2004probing} and the findings of Ansari et al.~\cite{ansari2022water}. Moreover, achieving the same converged result in a significantly shorter simulation time further highlights the computational efficiency of the Lambda-ABF-OPES method.

To evaluate the robustness of the method, we performed a detailed convergence test by analyzing the convergence time of $\Delta G$ and comparing the results with those obtained using Lambda-ABF alone. The convergence time was defined as the point from which all subsequent data points remain within the specified tolerance of the mean, providing a clear metric for determining when the simulation data stabilizes. A tolerance of 0.2~kcal/mol was used in this analysis, representing just 20\% of the commonly accepted convergence threshold for binding free energy calculations. This strict criterion underscores the precision and reliability of the method.

\begin{figure}[H]
	\centering
	\includegraphics[width=1.0\textwidth]{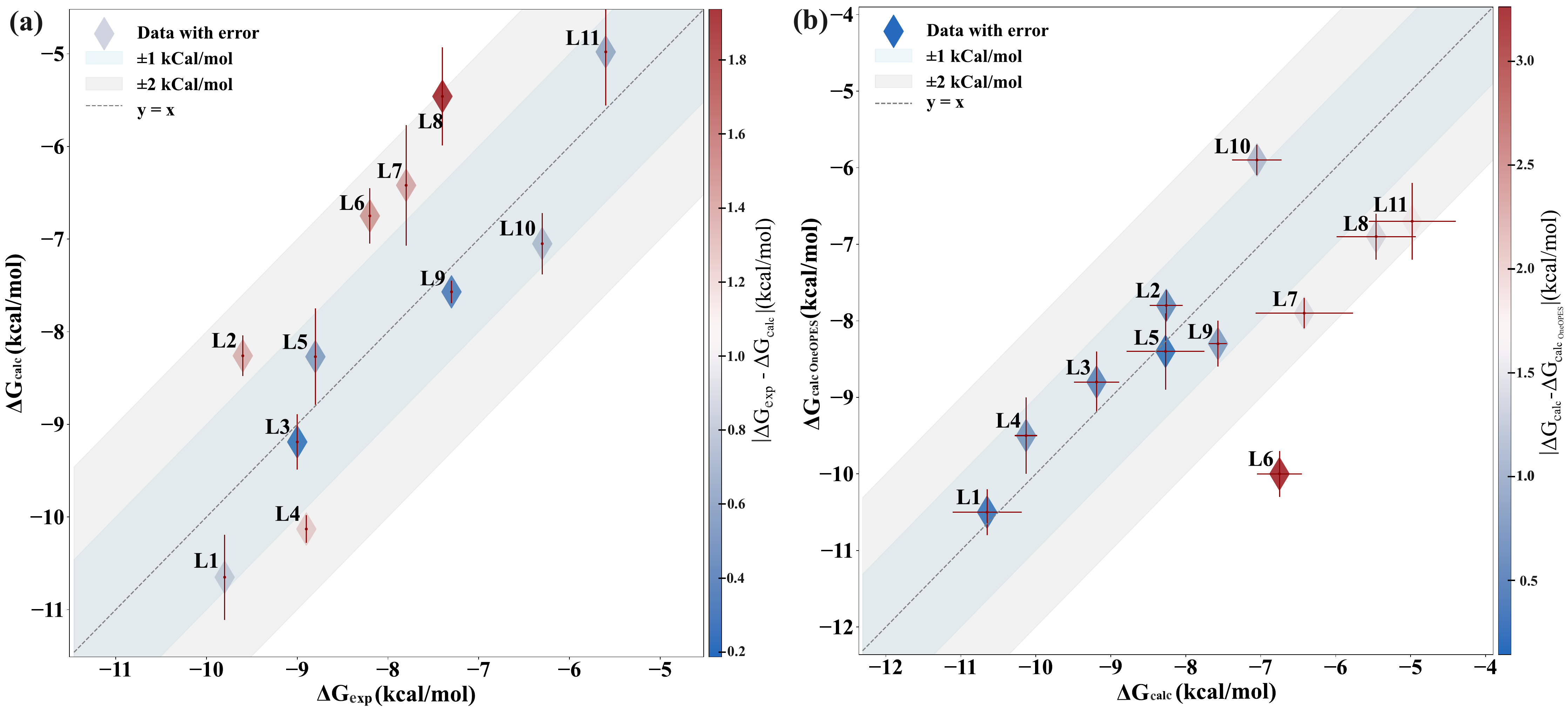}   
\caption{\textbf{Experimental vs. Calculated $\Delta G$ Values.} (a) Calculated $\Delta G$ values and their associated errors represent the mean and standard error from three independent repeats for each ligand. The dark shaded region spans $\pm$1~kcal/mol, while the lighter region spans $\pm$2~kcal/mol. The color bar indicates the absolute difference between experimental and computed values. Pearson’s $r$, RMSE, and MAE are 0.81, 1.10, and 0.90~kcal/mol, respectively. (b) Comparison of calculated $\Delta G$ values from this work with those from Ref.~\citenum{karrenbrock2024absolute}.} 
	\label{fig:fig3}
\end{figure}

\begin{figure}[H]
	\centering
	\includegraphics[width=0.9\textwidth]{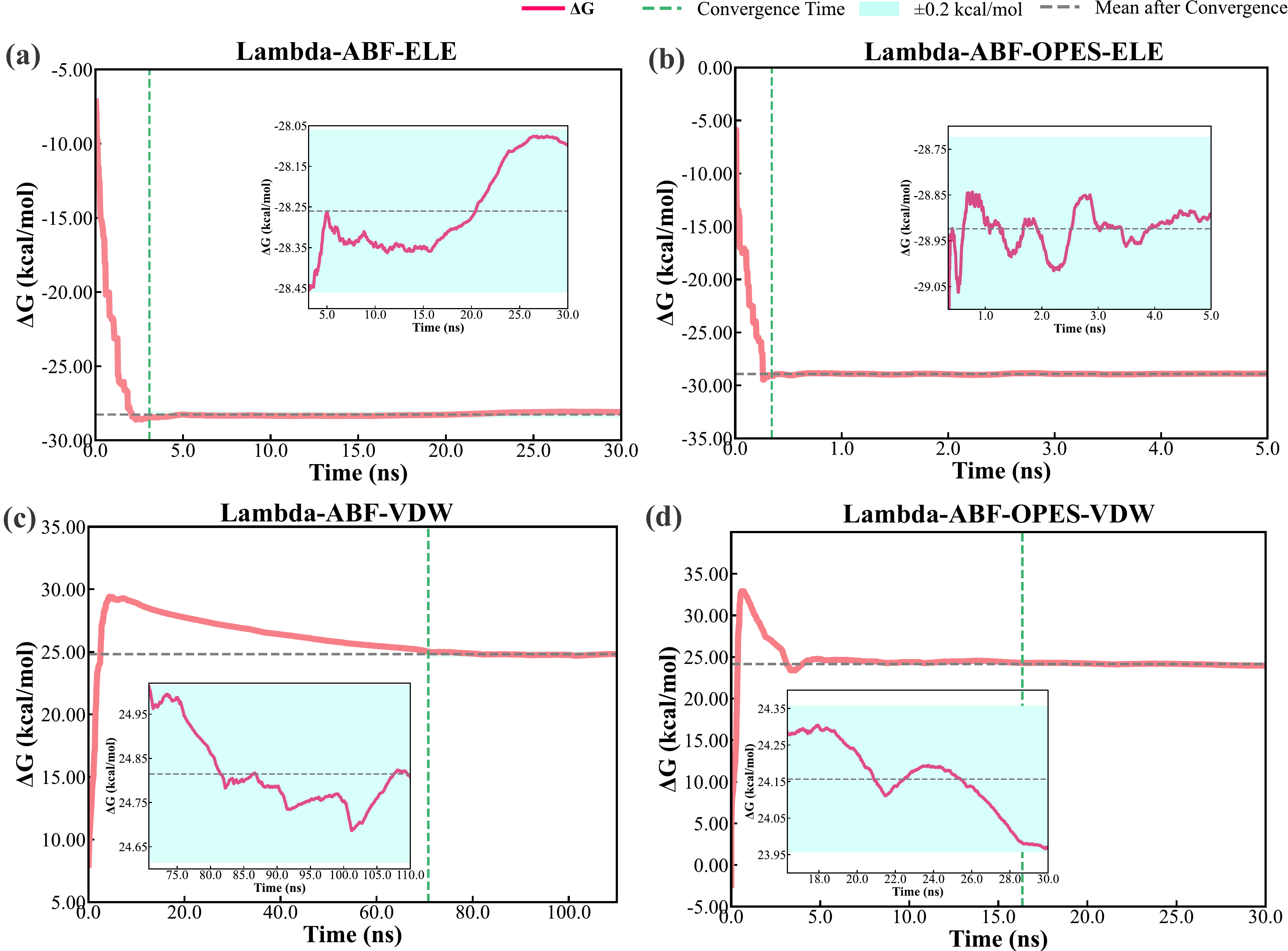}   
\caption{\textbf{$\Delta G$ convergence comparison between Lambda-ABF and Lambda-ABF-OPES }. The $\Delta G$ over time for the (a)-(b) ELE leg and the (c)-(d) VDW leg of ligand 8 complex phase, calculated using Lambda-ABF and Lambda-ABF-OPES methods, respectively. Each panel includes a zoomed-in plot of the converged area.}

	\label{fig:conv}
\end{figure}

Figure~\ref{fig:conv} illustrates an example of a $\Delta G$ convergence plot over time for the ELE and VDW legs of the ligand 8-BRD4 complex phase using the Lambda-ABF and Lambda-ABF-OPES methods in a single replica (see SI Fig.~S14-S18 for additional comparisons). In this case, the acceleration in convergence time is a factor of 9 for ELE and 4 for VDW. When considering different replicas, the convergence speedup for ELE ranges from 5 to 9 times, while for VDW, it ranges from 3 to 5 times.

The convergence analysis results for all ligands using Lambda-ABF-OPES are summarized in Table~\ref{tab:EXP_Cal}. The average convergence time for the complex phase was less than 3 ns for the ELE leg and approximately 20 ns for the VDW leg, highlighting the rapid stabilization of the $\Delta G$ in these components. In the solvent phase, the ELE leg showed convergence for most ligands within 1--2 ns or less, while the VDW leg converged within 2--3 ns. These results demonstrate a marked improvement in convergence speed compared to other currently used methods, which typically requires significantly longer simulation times to achieve comparable stability \cite{gapsys2021accurate,karrenbrock2024absolute,aldeghi2016accurate}

In practical applications, such efficiency gains can translate into a substantial reduction in computational costs, enabling more extensive exploration of chemical space or greater statistical sampling within the same resource constraints. The combination of simplicity, accuracy, and computational efficiency makes this approach a promising tool for the drug discovery field.

\begin{table}[htbp]
\centering
\scriptsize 
\begin{tabular}{ccccccccc}
\hline
\multicolumn{1}{c}{} & \multicolumn{1}{c}{} & \multicolumn{1}{c}{} & \multicolumn{1}{c}{} & \multicolumn{1}{c}{} & \multicolumn{4}{c}{\textbf{ns per walker}} \\ \cline{6-9} 
\multicolumn{1}{c}{} & \multicolumn{1}{c}{} & \multicolumn{1}{c}{} & \multicolumn{1}{c}{} & \multicolumn{1}{c}{} & \multicolumn{2}{c}{\textbf{Complex}}  & \multicolumn{2}{l}{\textbf{Solvent}} \\ \cline{6-9} 
\multicolumn{1}{c}{\multirow{-3}{*}{\textbf{Compound}}} & \multicolumn{1}{c}{\multirow{-3}{*}{$\mathbf{\Delta G_{exp}}$}} & \multicolumn{1}{c}{\multirow{-3}{*}{$\mathbf{\Delta G_{calc}}$}} & \multicolumn{1}{c}{\multirow{-3}{*}{$\mathbf{\Delta G_{calc}}$-$\mathbf{\Delta G_{exp}}$}} & \multicolumn{1}{c}{\multirow{-3}{*}{\textbf{PDB}}} & \textbf{ELE}           & \textbf{VDW}          & \textbf{ELE}          & \textbf{VDW}         \\ \hline
L1-BRD4  & -9.8 $\pm$ 0.1~\cite{ciceri2014dual}                     & -10.56 $\pm$ 0.46 & -0.85 & 4OGI  & 0.9 & 22.5 & 1.9 & 3.5\\
L2-BRD4  & -9.6 $\pm$ 0.1~\cite{filippakopoulos2010selective}       & -8.26  $\pm$ 0.22 & 1.34  & 3MXF  & 2.2 & 19.2 & 0.9 & 2.2\\
L3-BRD4  & -9.0 $\pm$ 0.1~\cite{picaud2013rvx}                      & -9.19  $\pm$ 0.30 & -0.19 & 4MR3  & 2.6 & 17.8 & 0.4 & 2.4\\
L4-BRD4  & -8.9 $\pm$ 0.1~\cite{ciceri2014dual}                     & -10.13 $\pm$ 0.15 & -1.23 & 4OGJ  & 2.6 & 24.8 & 2.7 & 3.5 \\
L5-BRD4  & -8.8 $\pm$ 0.1~\cite{hewings2013optimization}            & -8.27 $\pm$ 0.52 & 0.09   & 4J0R  & 2.1 & 15.9 & 0.9 & 2.7\\
L6-BRD4  & -8.2 $\pm$ 0.1~\cite{filippakopoulos2012benzodiazepines} & -6.75  $\pm$ 0.30 & 1.45  & 3U5L  & 0.6 & 20.2 & 0.5 & 3.5\\
L7-BRD4  & -7.8 $\pm$ 0.1~\cite{picaud2013rvx}                      & -6.42  $\pm$ 0.65 & 1.38  & 4MR4  & 0.9 & 20.0 & 1.4 & 3.4\\
L8-BRD4  & -7.4 $\pm$ 0.1~\cite{filippakopoulos2012benzodiazepines} & -5.46  $\pm$ 0.53 & 1.94  & 3U5J  & 0.7 & 20.7 & 0.4 & 2.5\\
L9-BRD4  & -7.3 $\pm$ 0.1~\cite{hewings2013optimization}            & -7.57  $\pm$ 0.12 & -0.27 & 3SVG  & 1.5 & 17.4 & 0.5 & 2.7\\
L10-BRD4 & -6.3 $\pm$ 0.0~\cite{fish2012identification}             & -7.05  $\pm$ 0.33 & -0.75 & 4HBV  & 1.8 & 14.8 & 0.2 & 1.6\\
L11-BRD4 & -5.6~\cite{vidler2013discovery}                          & -4.98  $\pm$ 0.58 & 0.62  & Model & 0.2 & 10.7 & 0.2 & 1.9\\ \hline
Benzamidine-Trypsin & -6.36~\cite{talhout2003understanding,talhout2004probing}   & -6.2  $\pm$ 0.65 & 0.16  & 3ATL & 0.98 & 17.0 & 0.34 & 3.52\\ \hline
\end{tabular}
\caption{\textbf{Summary of the BRD4 and Trypsin binding free energy results using Lambda-ABF-OPES}. The experimental ($\Delta G_{\text{exp}}$) and calculated ($\Delta G_{\text{calc}}$) values for each ligand are presented. All $\Delta G$ values are reported in kcal/mol. The calculated $\Delta G_{\text{calc}}$ and associated errors represent the mean and standard error of the mean, derived from three replicates for each ligand. The PDB files used as input are listed. The average convergence time over three replicas for each ELE and VDW leg in the complex and solvent phases is reported in ns per replica. The total simulation time for the ELE and VDW legs in the complex phase is 5 ns and 30 ns per walker, respectively. For the solvent phase, each leg is run for 5 ns per walker.
}
\label{tab:EXP_Cal}
\end{table}


Accurately predicting the binding affinity between small molecules and their target proteins remains a critical challenge in drug discovery, with far-reaching implications for the speed and efficiency of therapeutic development. 

In this study, we introduced a new hybrid approach that combines Lambda-ABF with the exploratory version of the OPES method. This novel integration leverages the complementary strengths of ABF and OPES$_{e}$ to overcome critical limitations in alchemical free energy calculations, including inefficient exploration of configurational space and kinetic trapping in energy landscapes. By applying biases to the Lambda collective variable (CV) and incorporating the AMOEBA polarizable force field alongside DBC restraints, our method achieves unprecedented levels of sampling efficiency, up to nine times faster than the original Lambda-ABF technique.

Our application of this hybrid method to a diverse set of 11 drug-like molecules targeting BRD4 bromodomains and Benzamidine-Trypsin complex yielded close alignment between our computational results and experimental data. Importantly, this was achieved while using the highly accurate AMOEBA polarizable force field, demonstrating the feasibility of this approach for real-world drug discovery applications. This approach can be naturally extended to neural networks methodologies including Machine Learning Interatomic Potentials (MLIP) \cite{ple2023force,ple2024fennol} and Foundation models whose additional computational cost for free energy computations compared to FFs has limited to date their use in production.

By integrating state-of-the-art methodologies and harnessing their synergistic advantages, this work provides a robust tool for the rapid and reliable advancement of novel therapeutics. Though applied in an alchemical context in this study, this methodology shows promise for broader applicability within enhanced sampling techniques and lays the groundwork for its integration into a more general framework, which we plan to further investigate in future work.

\section*{Data Availability Statement}
The Tinker-HP and colvars input files used, and the structures
are available on GitHub at:  
\href{https://github.com/ansarinarjes/ABFE-L-ABF-OPES.git}{https://github.com/ansarinarjes/ABFE-L-ABF-OPES.git}

\section*{Conflict of interest/Competing interests} 
L. L., and J.-P. P. are co-founders and shareholders of Qubit Pharmaceuticals. The remaining authors declare no competing interests.

\section{Supporting Information}
The Supporting Information contains more detailed descriptions of the MD simulations and Lambda-ABF-OPES, including applied restraints, DBC selection, performance, and convergence analyses. In total, there are 3 tables and 23 figures with corresponding descriptions.

\section{Acknowledgment}
We thank Haochuan Chen for implementing OPES in the Colvars library, building on Michele Invernizzi's work. We also thank Chengwen Liu and Timoth\'e Melin (Qubit Pharmaceuticals) for their assistance with the parameterization of complex ligands. We thank the Grand Équipement de Calcul Intensif (GENCI), Institut du Développement et des Ressources en Informatique (IDRIS), and Centre Informatique de l’Enseignement Supérieur (CINES), France, for their support of this work through grant no. AD010715770 and A0160714153. This work has received funding from the European Research Council (ERC) under the European Union's Horizon 2020 research and innovation program (grant agreement No 810367), project EMC2 (JPP).

\section{Technical Appendix}
\subsection{Analytical lambda derivatives for the polarization energy using Particle Mesh Ewald}
A simple interpolation of polarization between the end states reads:
\[E_{pol}(\br,\lambda)=\lambda E_{pol}(\br,1)+(1-\lambda) E_{pol}(\br,0)\]
so that:
\[\frac{\partial E_{pol}}{\partial \lambda}(\br,\lambda)= E_{pol}(\br,1)- E_{pol}(\br,0)\]
which requires two resolutions of the polarization equations per timestep as stated in the main text.
Alternatively, let's consider the polarization energy associated to a scaling of the polarizabilities and the permanent multipoles of the "alchemical" part of the system, the complete system being made of N atoms:
\[E_{pol}(\br,\lambda,\bmu(\br,\lambda))=\frac{1}{2}\bmu \bT(\br,\lambda) \bmu -\bmu \bE(\br,\lambda)\] where $\bT$ is the (3N,3N) polarization matrix and $\bE$ the 3N vector of the permanent electric fields on the polarizable sites. The Hellman-Feynman theorem yields:
\[\frac{d E_{pol}}{d \lambda}=\frac{1}{2}\bmu \frac{\partial \bT}{\partial \lambda} \bmu -\bmu \frac{\partial \bE}{\partial \lambda}\]
The derivative contribution due to the $\bT$ is trivial to compute because it is only associated to its diagonal part $\balpha^{-1}(\lambda)$, with $\balpha$ collecting the diagonal polarizability tensors. In the context of periodic boundary conditions computed with Particle Mesh Ewald, the second one can be separated in 3 given the various component of $\frac{\partial \bE}{\partial \lambda}=\frac{\partial\bE_{real}}{\partial \lambda}+\frac{\partial \bE_{self}}{\partial \lambda}+\frac{\partial \bE_{recip}}{\lambda}$. The first two terms can be directly computed.
If only fixed charge are used (and no permanent multipoles) then  he last one can be reformulated as:
\[\frac{\partial }{\partial \lambda}(\bmu  \bE_{recip}(\lambda))=\frac{\partial }{\partial \lambda}(\bQ(\lambda) \bV_{recip}(\bmu))=\frac{\partial }{\partial \lambda}(\bQ(\lambda)) \bV_{recip}(\bmu)\]
where $\bQ(\lambda)$ is the vector (of size N) containing the permanent charges and $\bV_{recip}$ the vector of same size containing the reciprocal potential due to the induced dipoles. This potential is readily available to compute the permanent polarization energy and $\frac{\partial \bQ(\lambda)}{\partial \lambda}$ is trivial to compute. The same reasoning is naturally extended to permanent multipoles of higher order by involving derivatives of $V_{recip}(\bmu)$ which are always available to compute the permanent polarization energy.

\bibliography{main}
\end{document}